\documentclass{svjour2}                    
\smartqed  
\usepackage{graphicx}
\usepackage{amssymb,amsmath,epsfig,graphicx}
\def\av#1{\left\langle#1\right\rangle}
\def\di{\partial}
\def\to{\rightarrow}
\def\Ka{K_{\alpha}}
\def\sKa{\sqrt{K_{\alpha}}}

\journalname{Journal of Statistical Physics}
\begin{document}

\title{On distributions of functionals of anomalous diffusion paths}

\titlerunning{Functionals of anomalous diffusion paths}        

\author{Shai Carmi \and Lior Turgeman \and Eli Barkai}


\institute{Shai Carmi \email{scarmi@shoshi.ph.biu.ac.il} \and Lior
Turgeman \and Eli Barkai \email{barkaie@mail.biu.ac.il} \at
Department of Physics and Advanced Materials and Nanotechnology Institute, Bar-Ilan University, Ramat Gan 52900, Israel}

\date{\today}

\maketitle

\begin{abstract}
Functionals of Brownian motion have diverse applications in physics,
mathematics, and other fields. The probability density function
(PDF) of Brownian functionals satisfies the Feynman-Kac formula,
which is a Schr\"{o}dinger equation in imaginary time. In recent
years there is a growing interest in particular functionals of
non-Brownian motion, or anomalous diffusion, but no equation existed
for their PDF. Here, we derive a \emph{fractional} generalization of
the Feynman-Kac equation for functionals of anomalous paths based on
sub-diffusive continuous-time random walk. We also derive a backward
equation and a generalization to L\'{e}vy flights. Solutions are
presented for a wide number of applications including the occupation
time in half space and in an interval, the first passage time, the
maximal displacement, and the hitting probability. We briefly
discuss other fractional Schr\"{o}dinger equations that recently
appeared in the literature.


\end{abstract}

\section{Introduction}

A Brownian functional is defined as $A=\int_0^tU[x(\tau)]d\tau$,
where $x(t)$ is a trajectory of a Brownian particle and $U(x)$ is a
prescribed function \cite{MajumdarReview}. Functionals of diffusive
motion arise in numerous problems across a variety of scientific
fields from condensed matter physics
\cite{ComtetReview,KPZRandomwalk,ProteinPulling}, to hydrodynamics
\cite{FriedrichPLA06}, meteorology \cite{Weather}, and finance
\cite{ExponentialFunctionals,Economy}. The distribution of these
functionals satisfies a Schr\"{o}dinger-like equation, derived in
1949 by Kac inspired by Feynman's path integrals \cite{Kac1949}.
Denote by $G(x,A,t)$ the joint probability density function (PDF) of
finding, at time $t$, the particle at $x$ and the functional at $A$.
The Feynman-Kac theory asserts that (for $U(x)>0$)
\cite{MajumdarReview,Kac1949}
\begin{equation}
\label{Feynman-Kac} \frac{\di}{\di t}G(x,p,t)=K\frac{\di^2}{\di
x^2}G(x,p,t)-pU(x)G(x,p,t),
\end{equation}
where the equation is in Laplace space, $A\to p$, and $K$ is the
diffusion coefficient.

The celebrated Feynman-Kac equation \eqref{Feynman-Kac} describes
functionals of normal Brownian motion. However, we know today that
in a vast number of systems the underlying processes exhibit
anomalous, non-Brownian sub-diffusion, as reflected by the nonlinear
relation: $\av{x^2}\sim t^{\alpha}\;,\;0<\alpha<1$
\cite{Havlin,Bouchaud,KlafterReview2000,AnomalousTransportBook,Mandelbrot}.
While a few specific functionals of anomalous paths have been
investigated \cite{OccTimeMajumdarPRL,BarkaiJSP06}, a general theory
is still missing.

Several functionals of anomalous diffusion are of interest. For
example, the time spent by a particle in a given domain, or the
occupation time, is given by the functional
$A=\int_0^tU[x(\tau)]d\tau$, where $U(x)=1$ in the domain and is
zero otherwise
\cite{WeissResidence,KlafterResidence,AgmonResidence,AgmonResidenceMoments}.
Such a functional can be used in kinetic studies of chemical
reactions that take place exclusively in the domain. Consider for
example a particle diffusing in a medium containing an interval that
is absorbing at rate $R$. The average survival probability of the
particle is $\av{\exp(-RA)}$ \cite{GrebenkovPRE}. Two other related
functionals are the occupation time in the positive half-space
($U(x)=\Theta(x)$) and the local time ($U(x)=\delta(x)$)
\cite{OccTimeMajumdarPRL,BarkaiJSP06,Lamperti,OccTimeMajumdarPRE,KarlinTaylor}.

Another interesting family of functionals arises in the study of NMR
\cite{NMRReview}. In a typical NMR experiment, the macroscopic
measured signal can be written as $E=\av{e^{i\varphi}}$ where
$\varphi = \gamma \int_0^t B[x(\tau)]d\tau$ is the phase accumulated
by each spin, $\gamma$ is the gyromagnetic ratio, $B(x)$ is a
spatially-inhomogeneous external magnetic field, and $x(\tau)$ is
the trajectory of each particle. NMR therefore indirectly encodes
information regarding the motion of the particles. Common choices of
the magnetic field $B$ are $B(x)=x$ and $B(x)=x^2$ \cite{NMRReview}.
For dispersive systems with inhomogeneous disorder where the motion
of the particles is non-Brownian, the phase $\varphi$ is a
non-Brownian functional with $U(x)=x$ or $U(x)=x^2$.

In this paper, we develop a general theory of non-Brownian
functionals. The process we consider as the mechanism that leads to
non-Brownian transport is the sub-diffusive continuous-time
random-walk (CTRW). This is an important and widely investigated
process that is frequently used to describe the motion of particles
in disordered systems
\cite{Havlin,Bouchaud,KlafterReview2000,MontrollWeiss,ScherMontroll}.
In the scaling limit of this process, we derive the following
\emph{fractional} Feynman-Kac equation:
\begin{equation}
\label{forward_eq} \frac{\di}{\di t}G(x,p,t)=\Ka\frac{\di^2}{\di
x^2}{\cal D}_t^{1-\alpha}G(x,p,t)-pU(x)G(x,p,t),
\end{equation}
where the symbol ${\cal D}_t^{1-\alpha}$ is Friedrich's
\emph{substantial fractional derivative} and is equal in Laplace
space $t \to s$ to $[s+pU(x)]^{1-\alpha}$ \cite{FriedrichPRL06}. In
the rest of the paper, we derive Eq. \eqref{forward_eq} and its
backward version and then investigate applications for specific
functionals of interest. A brief report of part of the results has
recently appeared in \cite{BarkaiPRL09}.

\section{Derivation of the equations}
\label{derivation_sect}

We use the continuous-time random-walk (CTRW) model as the
underlying process leading to anomalous diffusion
\cite{Havlin,Bouchaud,KlafterReview2000,MontrollWeiss,ScherMontroll}.
In CTRW, an infinite one-dimensional lattice with spacing $a$ is
assumed, and allowed jumps are to nearest neighbors only and with
equal probability of jumping left or right. Waiting times between
jump events are independent identically distributed random variables
with PDF $\psi(\tau)$, and the process starts with a particle at
$x=x_0$. The particle waits at $x_0$ for time $\tau$ drawn from
$\psi(\tau)$ and then jumps with probability $1/2$ to either $x_0+a$
or $x_0-a$, after which the process is renewed. We assume that no
external forces are applied and that for long waiting times,
$\psi(\tau)\sim B_{\alpha}\tau^{-(1+\alpha)}/|\Gamma(-\alpha)|$. For
$0<\alpha<1$, the average waiting time is infinite and the process
is sub-diffusive with $\av{x^2}= 2\Ka t^{\alpha}/\Gamma(1+\alpha)$
($\Ka=a^2/(2B_{\alpha})$, units
$\textrm{m}^2/\textrm{sec}^{\alpha}$) \cite{BarkaiPRE00}. We look
for the differential equation that describes the distribution of
functionals in the scaling limit of this model.

\subsection{Derivation of the fractional Feynman-Kac equation}

\label{forward_section}

Recall that the functional is defined as $A=\int_0^tU[x(\tau)]d\tau$
and that $G(x,A,t)$ is the joint PDF of $x$ and $A$ at time $t$. For
the particle to be at $(x,A)$ at time $t$, it must have been at
$[x,A-\tau U(x)]$ at the time $t-\tau$ when the last jump was made.
Let $Q_n(x,A,t)dt$ be the probability of the particle to make its
$n$th jump into $(x,A)$ in the time interval $[t,t+dt]$. Thus,
\begin{equation}
\label{G_Q} G(x,A,t)=\int_0^tW(\tau)\sum_{n=0}^{\infty}Q_n[x,A-\tau
U(x),t-\tau]d\tau,
\end{equation}
where $W(\tau)=1-\int_0^{\tau}\psi(\tau')d\tau'$ is the probability
for \emph{not} moving in a time interval of length $\tau$.

To arrive into $(x,A)$ after $n+1$ jumps, the particle must have
arrived after $n$ jumps into either $[x-a,A-\tau U(x-a)]$ or
$[x+a,A-\tau U(x+a)]$, where $\tau$ the time between the jumps.
Since the probabilities of jumping left and right are equal, we can
write a recursion relation for $Q_n$:
\begin{align}
\label{Q_recursion} Q_{n+1}(x,A,t)=\int_0^t\psi(\tau)
&\left\{\frac{1}{2}Q_n[x+a,A-\tau U(x+a),t-\tau]\right. \\
+&~~\left.\frac{1}{2}Q_n[x-a,A-\tau
U(x-a),t-\tau]\right\}d\tau\nonumber,
\end{align}
where $\psi(\tau)$ is the PDF of $\tau$, the time between jumps. For
$n=0$ (no jumps were made), $Q_0=\delta(x-x_0)\delta(A)\delta(t)$.

Assume that $U(x)\geq 0$ for all $x$ and thus $A\geq 0$ (an
assumption we will relax later). Let $Q_n(x,p,t)$ be the Laplace
transform $A\to p$ of $Q_n(x,A,t)$--- we use along this work the
convention that the variables in parenthesis define the space we are
working in. We note that
\begin{align*}
\int_0^{\infty}e^{-pA}Q_n[x,A-\tau U(x),t]dA&= e^{-p\tau
U(x)}\int_{0}^{\infty}e^{-pA'}Q_n(x,A',t)dA'\nonumber \\&=e^{-p\tau
U(x)}Q_n(x,p,t),
\end{align*}
where we used the fact that $Q_n(x,A,t)=0$ for $A<0$. Thus, Laplace
transforming $A\to p$ Eq. \eqref{Q_recursion} we find
\begin{align}
\label{Q_recursion_Ap}
Q_{n+1}(x,p,t)&=\frac{1}{2}\int_0^t\psi(\tau)e^{-p\tau
U(x+a)}Q_n(x+a,p,t-\tau)d\tau \nonumber \\
&+\frac{1}{2}\int_0^t\psi(\tau)e^{-p\tau
U(x-a)}Q_n(x-a,p,t-\tau)d\tau.
\end{align}
Laplace transforming $t\to s$ Eq. \eqref{Q_recursion_Ap} using the
convolution theorem,
\begin{align}
\label{Q_recursion_ts}
Q_{n+1}(x,p,s)&=\frac{1}{2}\hat\psi[s+pU(x+a)]Q_n(x+a,p,s) \nonumber \\
&+\frac{1}{2}\hat\psi[s+pU(x-a)]Q_n(x-a,p,s),
\end{align}
where $\hat\psi(s)$ is the Laplace transform of the waiting time
PDF. Fourier transforming $x\to k$ Eq. \eqref{Q_recursion_ts},
\begin{equation*}
Q_{n+1}(k,p,s)=\cos(ka)\int_{-\infty}^{\infty}e^{ikx}\hat\psi[s+pU(x)]Q_n(x,p,s)dx.
\end{equation*}
Applying the Fourier transform identity ${\cal
F}\{xf(x)\}=-i\frac{\di}{\di k}f(k)$,
\begin{equation}
\label{Q_recursion_kps}
Q_{n+1}(k,p,s)=\cos(ka)\hat\psi\left[s+pU\left(-i\frac{\di}{\di
k}\right)\right]Q_n(k,p,s).
\end{equation}
Note that the order of the terms is important:
$\hat\psi\left[s+pU\left(-i\frac{\di}{\di k}\right)\right]$ does not
commute with $\cos(ka)$. Summing Eq. \eqref{Q_recursion_kps} over
all $n$, using the initial condition $Q_0(k,p,s)=e^{ikx_0}$, and
rearranging, we obtain,
\begin{equation}
\label{Q_solution}
\sum_{n=0}^{\infty}Q_n(k,p,s)=\left\{1-\cos(ka)\hat\psi\left[s+pU\left(-i\frac{\di}{\di
k}\right)\right]\right\}^{-1}e^{ikx_0}.
\end{equation}

We next use our expression for $\sum_{n=0}^{\infty} Q_n$ to
calculate $G(x,A,t)$. Transforming Eq. \eqref{G_Q} $(x,A,t)\to
(k,p,s)$,
\begin{equation}
\label{G_Q_kps} G(k,p,s)=
\frac{1-\hat\psi\left[s+pU\left(-i\frac{\di}{\di
k}\right)\right]}{s+pU\left(-i\frac{\di}{\di
k}\right)}\sum_{n=0}^{\infty}Q_n(k,p,s),
\end{equation}
where we used $\hat
W(s)=\int_0^{\infty}e^{-st}\left[1-\int_0^t\psi(\tau)d\tau\right]dt=[1-\hat\psi(s)]/s$.
Substituting Eq. \eqref{Q_solution} into Eq. \eqref{G_Q_kps}, we
find the formal solution
\begin{align}
\label{P_usk} G(k,p,s)&=
\frac{1-\hat\psi\left[s+pU\left(-i\frac{\di}{\di
k}\right)\right]}{s+pU\left(-i\frac{\di}{\di k}\right)}
\times\nonumber \\ &\times
\left\{1-\cos(ka)\hat\psi\left[s+pU\left(-i\frac{\di}{\di
k}\right)\right]\right\}^{-1}e^{ikx_0}.
\end{align}

To derive a differential equation for $G(x,p,t)$, we recall the
waiting time distribution is $\psi(t) \sim
B_{\alpha}t^{-(1+\alpha)}/|\Gamma(-\alpha)|$ and write its Laplace
transform $\hat\psi(s)$ for $s\to 0$ as \cite{KlafterReview2000}
\begin{equation}
\label{psi_s_small} \hat\psi(s) \sim
1-B_{\alpha}s^{\alpha}\quad;\quad 0<\alpha<1\;,\;s\to 0.
\end{equation}
Substituting Eq. \eqref{psi_s_small} into Eq. \eqref{P_usk},
applying the small $k$ expansion $\cos(ka) \sim 1-k^2a^2/2$, and
neglecting the high order terms, we have
\begin{align*}
G(k,p,s)= \left[s+pU\left(-i\frac{\di}{\di
k}\right)\right]^{\alpha-1}\left\{\Ka k^2+\left[s+pU\left(-i\frac{\di}{\di
k}\right)\right]^{\alpha}\right\}^{-1}e^{ikx_0},
\end{align*}
where we used the generalized diffusion coefficient $\Ka\equiv
\lim_{a^2,B_{\alpha}\to 0}a^2/(2B_{\alpha})$ \cite{BarkaiPRE00}. By
neglecting the high order terms in $s$ and $k$ we effectively reach
the scaling limit of the lattice walk
\cite{Meer,Kotulski,HavlinJSP}. Rearranging the expression in the
last equation we find
\begin{align*}
sG(k,p,s)-e^{ikx_0}=&-\Ka k^2\left[s+pU\left(-i\frac{\di}{\di
k}\right)\right]^{1-\alpha}G(k,p,s)\nonumber
\\ &-pU\left(-i\frac{\di}{\di k}\right)G(k,p,s).
\end{align*}
Inverting $k\to x, s\to t$ we finally obtain our fractional
Feynman-Kac equation
\begin{equation}
\label{forward_eq_derive} \frac{\di}{\di
t}G(x,p,t)=\Ka\frac{\di^2}{\di x^2}{\cal
D}_t^{1-\alpha}G(x,p,t)-pU(x)G(x,p,t).
\end{equation}
The initial condition is $G(x,A,t=0)=\delta(x-x_0)\delta(A)$, or
$G(x,p,t=0)=\delta(x-x_0)$.

${\cal D}_t^{1-\alpha}$ is the fractional substantial derivative
operator introduced in \cite{FriedrichPRL06}:
\begin{equation}
\label{substantial_def_s}
{\cal D}_t^{1-\alpha}G(x,p,s)=[s+pU(x)]^{1-\alpha}G(x,p,s).
\end{equation}
In $t$ space,
\begin{equation}
\label{substantial_def} {\cal
D}_t^{1-\alpha}G(x,p,t)=\frac{1}{\Gamma(\alpha)}\left[\frac{\di}{\di
t}+pU(x)\right]\int_0^t\frac{e^{-(t-\tau)pU(x)}}{(t-\tau)^{1-\alpha}}G(x,p,\tau)d\tau.
\end{equation}
Thus, due to the long waiting times, the evolution of $G(x,p,t)$ is
non-Markovian and depends on the entire history.

In $s$ space, the fractional Feynman-Kac equation reads
\begin{equation}
\label{forward_eq_ps}
sG(x,p,s)-\delta(x-x_0)=\Ka\frac{\di^2}{\di
x^2}[s+pU(x)]^{1-\alpha}G(x,p,s)-pU(x)G(x,p,s).
\end{equation}

A few remarks should be made.

(\emph{i}) \emph{The integer Feynman-Kac equation.---} As expected,
for $\alpha=1$ our fractional equation \eqref{forward_eq_derive}
reduces to the (integer) Feynman-Kac equation \eqref{Feynman-Kac}.

(\emph{ii}) \emph{The fractional diffusion equation.---} For $p=0$,
$G(x,p=0,t)=\int_0^{\infty}G(x,A,t)dA$ reduces to $G(x,t)$, the
marginal PDF of finding the particle at $x$ at time $t$ regardless
of the value of $A$. Correspondingly, Eq. \eqref{P_usk} reduces to
the well-known Montroll-Weiss CTRW equation (for $x_0=0$)
\cite{KlafterReview2000,MontrollWeiss}:
\begin{equation*}
G(k,p=0,s)=\frac{1-\hat\psi(s)}{s}\frac{1}{1-\cos(ka)\hat\psi(s)}.
\end{equation*}
Eq. \eqref{forward_eq_derive} reduces to the fractional diffusion
equation:
\begin{equation}
\label{fractional_diffusion} \frac{\di}{\di
t}G(x,t)=\Ka\frac{\di^2}{\di x^2}{\cal
D}_{\textrm{RL},t}^{1-\alpha}G(x,t),
\end{equation}
where ${\cal D}_{\textrm{RL},t}^{1-\alpha}$ is the Riemann-Liouville
fractional derivative operator (${\cal
D}_{\textrm{RL},t}^{1-\alpha}G(x,s)\to s^{1-\alpha}G(x,s)$ in
Laplace $t\to s$ space) \cite{KlafterReview2000,SchneiderWyss}.


(\emph{iii}) \emph{The scaling limit.---} To derive our main
result--- the differential equation \eqref{forward_eq_derive}--- we
used the scaling, or continuum, limit to CTRW
\cite{BarkaiPRE00,Meer,Kotulski,HavlinJSP}. In this limit, we take
$a\to 0$ and $B_{\alpha}\to 0$, but keep $\Ka=a^2/(2B_{\alpha})$
finite. Recently, trajectories of this process were shown to obey a
certain class of stochastic Langevin equations
\cite{Fogedby,MagdziarzSimulations1,FriedrichSimulations}, hence
giving these paths a mathematical meaning.

(\emph{iv}) \emph{How to solve the fractional Feynman-Kac
equation.---} To obtain the PDF of a functional $A$, the following
recipe could be followed \cite{MajumdarReview}:
\begin{enumerate}
\item Solve Eq. \eqref{forward_eq_ps}, the fractional Feynman-Kac equation in $(x,p,s)$ space.
Eq. \eqref{forward_eq_ps} is a second order, ordinary differential equation in $x$.
\item Integrate the solution over all $x$
to eliminate the dependence on the final position of the particle.
\item Invert the solution $(p,s)\to (A,t)$, to obtain $G(A,t)$, the
PDF of $A$ at time $t$.
\end{enumerate}
We will later see (Section \ref{backward_section}) that the second
step can be circumvented by using a backward equation.

(\emph{v}) \emph{A general functional.---} When the functional is
not necessarily positive, the Laplace transform $A\to p$ must be
replaced by a Fourier transform. We show in the Appendix that in
this case the fractional Feynman-Kac equation looks like
\eqref{forward_eq_derive}, but with $p$ replaced by $-ip$,
\begin{equation}
\label{forward_eq_fourier} \frac{\di}{\di
t}G(x,p,t)=\Ka\frac{\di^2}{\di x^2}{\cal
D}_t^{1-\alpha}G(x,p,t)+ipU(x)G(x,p,t),
\end{equation}
where $G(x,p,t)$ is the Fourier transform $A\to p$ of $G(x,A,t)$
and ${\cal D}_t^{1-\alpha}\to[s-ipU(x)]^{1-\alpha}$ in Laplace $s$ space.

(\emph{vi}) \emph{L\'{e}vy flights.---} Consider CTRW with
displacements $\Delta_x$ distributed according to a symmetric PDF
$f(\Delta_x)\sim |\Delta_x|^{-(1+\mu)}$, with $0<\mu<2$. For this
distribution, the characteristic function is $f(k) \sim
1-C_{\mu}|k|^{\mu}$ \cite{KlafterReview2000}. This process is known
as a L\'{e}vy flight, and as we show in the Appendix, the fractional
Feynman-Kac equation for this case is (for $A\geq 0$)
\begin{equation}
\label{forward_eq_general} \frac{\di}{\di
t}G(x,p,t)=K_{\alpha,\mu}\nabla_x^{\mu}{\cal
D}_t^{1-\alpha}G(x,p,t)-pU(x)G(x,p,t),
\end{equation}
where $K_{\alpha,\mu}=C_{\mu}/B_{\alpha}$ (units
$\textrm{m}^{\mu}/\textrm{sec}^{\alpha}$), and ${\cal
D}_t^{1-\alpha}$ is the substantial fractional derivative operator
defined above (Eqs.
\eqref{substantial_def_s},\eqref{substantial_def}). $\nabla_x^{\mu}$
is the Riesz spatial fractional derivative operator defined in
Fourier $x\to k$ space as $\nabla_x^{\mu}\to -|k|^{\mu}$
\cite{KlafterReview2000}.

\subsection{A backward equation}
\label{backward_section}

In many cases we are only interested in the distribution of the
functional, $A$, regardless of the final position of the particle,
$x$. Therefore, it turns out quite convenient (see Section
\ref{Applications}) to obtain an equation for $G_{x_0}(A,t)$--- the
PDF of $A$ at time $t$, given that the process has started at $x_0$.

According to the CTRW model, the particle, after its first jump at
time $\tau$, is at either $x_0-a$ or $x_0+a$. Alternatively, the
particle does not move at all during the measurement time $[0,t]$.
Hence,
\begin{align}
\label{backward_recursion} G_{x_0}(A,t) = \int_0^t
\psi(\tau)&\left\{\frac{1}{2}G_{x_0+a}[A-\tau U(x_0),t-\tau]\right. \nonumber \\
+&~~\left.\frac{1}{2}G_{x_0-a}[A-\tau U(x_0),t-\tau]\right\}d\tau +
W(t)\delta[A-tU(x_0)].
\end{align}
Here, $\tau U(x_0)$ is the contribution to $A$ from the pausing time
at $x_0$ in the time interval $[0,\tau]$. The last term on the right hand side
of Eq. \eqref{backward_recursion} describes a motionless particle,
for which $A(t)=tU(x_0)$. We now Laplace transform Eq.
\eqref{backward_recursion} with respect to $A$ and $t$, using
techniques similar to those used in the previous subsection. This leads
to (for $A\geq 0$)
\begin{align*}
G_{x_0}(p,s) &=
\frac{1}{2}\hat\psi[s+pU(x_0)]\left[G_{x_0+a}(p,s)+G_{x_0-a}(p,s)\right]
\nonumber \\ &+ \frac{1-\hat\psi[s+pU(x_0)]}{s+pU(x_0)}.
\end{align*}
Fourier transform $x_0\to k_0$ of the last equation results in
\begin{align*}
G_{k_0}(p,s) &= \hat\psi\left[s+pU\left(-i\frac{\di}{\di
k_0}\right)\right]\cos(k_0a)G_{k_0}(p,s)
\nonumber \\
&+\frac{1-\hat\psi\left[s+pU\left(-i\frac{\di}{\di
k_0}\right)\right]}{s+pU\left(-i\frac{\di}{\di
k_0}\right)}\delta(k_0).
\end{align*}
As before, writing $\hat\psi(s)\sim 1-B_{\alpha}s^{\alpha}$ and
$\cos(k_0a)\sim 1-a^2k_0^2/2$, we have
\begin{align*}
\left[s+pU\left(-i\frac{\di}{\di
k_0}\right)\right]^{\alpha}G_{k_0}(p,s)&+\Ka{k_0}^2G_{k_0}(p,s)
\nonumber \\ &= \left[s+pU\left(-i\frac{\di}{\di
k_0}\right)\right]^{\alpha-1}\delta(k_0),
\end{align*}
where we used the generalized diffusion coefficient
$\Ka=a^2/(2B_{\alpha})$. Operating on both sides with
$\left[s+pU\left(-i\frac{\di}{\di k_0}\right)\right]^{1-\alpha}$,
\begin{align*}
sG_{k_0}(p,s)-\delta(k_0) =
&-\Ka\left[s+pU\left(-i\frac{\di}{\di
k_0}\right)\right]^{1-\alpha}{k_0}^2G_{k_0}(p,s)\nonumber
\\ &-pU\left(-i\frac{\di}{\di k_0}\right)G_{k_0}(p,s).
\end{align*}
Inverting $k_0\to x_0$ and $s\to t$, we obtain the backward
fractional Feynman-Kac equation:
\begin{equation}
\label{backward_eq} \frac{\di}{\di t}G_{x_0}(p,t)=\Ka{\cal
D}_t^{1-\alpha}\frac{\di^2}{\di
x_0^2}G_{x_0}(p,t)-pU(x_0)G_{x_0}(p,t).
\end{equation}
Here, ${\cal D}_t^{1-\alpha}$ equals in Laplace $t\to s$ space
$[s+pU(x_0)]^{1-\alpha}$. The initial condition is
$G_{x_0}(A,t=0)=\delta(A)$, or $G_{x_0}(p,t=0)=1$. In Eq.
\eqref{forward_eq_derive} the operators depend on $x$ while in Eq.
\eqref{backward_eq} they depend on $x_0$. Therefore, Eq.
\eqref{forward_eq_derive} is called the forward equation while Eq.
\eqref{backward_eq} is called the backward equation. Notice that
here, the fractional derivative operator appears to the left of the
Laplacian $\di^2/\di {x_0}^2$, in contrast to the forward equation
\eqref{forward_eq_derive}.

In the general case when the functional is not necessarily positive
and jumps are distributed according to a symmetric PDF
$f(\Delta_x)\sim |\Delta_x|^{-(1+\mu)}$, $0<\mu<2$, the backward equation
becomes (see the Appendix)
\begin{equation}
\label{general_backward_eq} \frac{\di}{\di t}G_{x_0}(p,t) =
K_{\alpha,\mu}{\cal
D}_t^{1-\alpha}\nabla_{x_0}^{\mu}G_{x_0}(p,t)+ipU(x_0)G_{x_0}(p,t).
\end{equation}
Here, $p$ is the Fourier pair of $A$, ${\cal D}_t^{1-\alpha}\to
[s-ipU(x_0)]^{1-\alpha}$ in Laplace $t\to s$ space, and
$\nabla_{x_0}^{\mu}\to -|k_0|^{\mu}$ in Fourier $x_0\to k_0$ space
(see also comments (\emph{v}) and (\emph{vi}) at the end of section
\ref{forward_section} above).

\section{Applications}
\label{Applications}

In this section, we describe a number of ways by which our equations
can be solved to obtain the distribution, the moments, and other
properties of functionals of interest.

\subsection{Occupation time in half-space}
\label{occupation_section}

\begin{figure}[t]
\begin{center}
\hspace{-0.5cm}
\epsfig{file=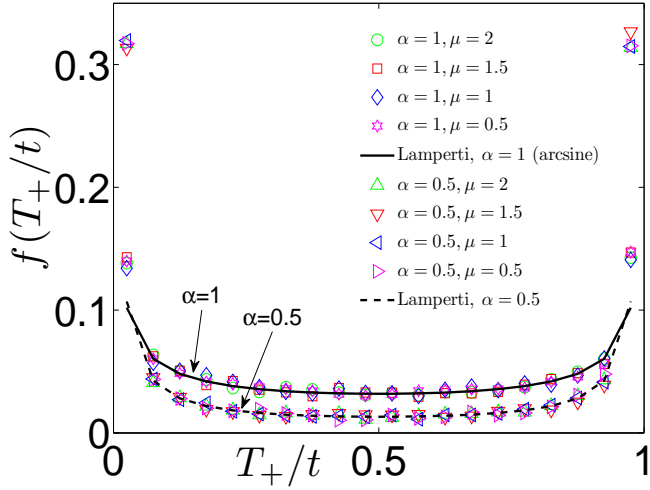,height=7cm,width=9cm}
\vspace{-0.5cm} \end{center} \caption{The PDF of the occupation
fractions in half-space $T_+/t$. Trajectories of diffusing particles
were generated using the methods of
\cite{MagdziarzSimulations1,FriedrichSimulations,MagdziarzSimulations2}
with parameter values $\Delta \tau=\overline{\Delta \tau}=10^{-3}$,
$\Delta t=10^{-2}$ (as defined in \cite{MagdziarzSimulations1}),
$\Ka=1$, and $x_0=0$. Simulations ended at $t=10^4$ and included
$10^4$ trajectories. Simulation results for $\alpha=0.5,1$ and
$\mu=0.5,1,1.5,2$ (see Eq. \eqref{general_backward_eq}) are shown as
symbols (see legend). Theoretical curves correspond to Lamperti's
PDF, Eq. \eqref{Lamperti_eq} (the arcsine distribution for
$\alpha=1$), and are plotted as a solid line for $\alpha=1$ and as a
dashed line for $\alpha=0.5$. It can be seen that the distribution
of occupation fractions is determined by $\alpha$ but not by $\mu$.}
\label{fig_occ_levy}
\end{figure}

Define the occupation time of a particle in the positive half-space
as $T_+=\int_0^t\Theta[x(\tau)]d\tau$ ($\Theta(x)=1$ for $x\geq 0$
and is zero otherwise)
\cite{OccTimeMajumdarPRL,BarkaiJSP06,OccTimeMajumdarPRE}. To find
the distribution of occupation times, we consider the backward
equation (\eqref{backward_eq}, transformed $t\to s$):
\begin{equation}
\label{occupation_ode} sG_{x_0}(p,s)-1=
\begin{cases}
\Ka s^{1-\alpha}\frac{\di^2}{\di
{x_0}^2}G_{x_0}(p,s)&x_0<0,\\
\Ka(s+p)^{1-\alpha}\frac{\di^2}{\di
{x_0}^2}G_{x_0}(p,s)-pG_{x_0}(p,s)&x_0>0.
\end{cases}
\end{equation}
These are second order, ordinary differential equations in $x_0$.
Solving the equations in each half-space separately, demanding that
$G_{x_0}(p,s)$ is finite for $|x_0| \to \infty$,
\begin{equation}
\label{occupation_ode_sol} G_{x_0}(p,s)=
\begin{cases}
C_0\exp(x_0s^{\alpha/2}/\sKa)+\frac{1}{s} & x_0<0, \\
C_1\exp[-x_0(s+p)^{\alpha/2}/\sKa]+\frac{1}{s+p} &
x_0>0.
\end{cases}
\end{equation}
For $x_0\to -\infty$, the particle is never at $x>0$ and thus
$G_{x_0}(T_+,t)=\delta(T_+)$ and $G_{x_0}(p,s)=s^{-1}$, in
accordance with Eq. \eqref{occupation_ode_sol}. Similarly, for
$x_0\to +\infty$, $G_{x_0}(T_+,t)=\delta(T_+-t)$ and
$G_{x_0}(p,s)=(s+p)^{-1}$. Demanding that $G_{x_0}(p,s)$ and its
first derivative are continuous at $x_0=0$, we obtain a pair of
equations for $C_0,C_1$:
\begin{equation*}
C_0+s^{-1}=C_1+(s+p)^{-1}\qquad;\qquad
C_0s^{\alpha/2}=-C_1(s+p)^{\alpha/2},
\end{equation*}
whose solution is
\begin{equation*}
C_0=-\frac{p(s+p)^{\alpha/2-1}}{s[s^{\alpha/2}+(s+p)^{\alpha/2}]}\quad;\quad
C_1=\frac{ps^{\alpha/2-1}}{(s+p)[s^{\alpha/2}+(s+p)^{\alpha/2}]}.
\end{equation*}
Assuming the process starts at $x_0=0$, $G_0(p,s)=C_1+(s+p)^{-1}$,
or, after some simplifications:
\begin{equation}
\label{occupation_Gps}
G_0(p,s)=
\frac{s^{\alpha/2-1}+(s+p)^{\alpha/2-1}}{s^{\alpha/2}+(s+p)^{\alpha/2}}.
\end{equation}
Using \cite{GodrecheLuck}, the PDF of $p_+\equiv T_+/t$, for long
times, is the (symmetric) \emph{Lamperti} PDF:
\begin{equation}
\label{Lamperti_eq}
f(p_+)=\frac{\sin(\pi\alpha/2)}{\pi}\frac{(p_+)^{\alpha/2-1}(1-p_+)^{\alpha/2-1}}{(p_+)^{\alpha}+(1-p_+)^{\alpha}+2(p_+)^{\alpha/2}(1-p_+)^{\alpha/2}\cos(\pi\alpha/2)}.
\end{equation}
This equation has been previously derived using different methods
\cite{BarkaiJSP06,Lamperti,GodrecheLuck,BouchaudOccTime} and was
also shown to describe occupation times of on and off states in
blinking quantum dots
\cite{MargolinPRL,MargolinJSP,StefaniPhysToday}. Naively, one
expects the particle to spend about half the time at $x>0$. In
contrast, we learn from Eq. \eqref{Lamperti_eq} that the particle
tends to spend most of the time at either $x>0$ or $x<0$: $f(p_+)$
has two peaks at $p_+=1$ and $p_+=0$ (Fig. \ref{fig_occ_levy}). This
is exacerbated in the limit $\alpha \to 0$, where the distribution
converges to two delta functions at $p_+=1$ and at $p_+=0$. For
$\alpha=1$ (Brownian motion) we recover the well-known arcsine law
of L\'{e}vy
\cite{MajumdarReview,OccTimeMajumdarPRL,BarkaiJSP06,Watanabe}.

We note that the PDF \eqref{Lamperti_eq} is a special case of the
more general, two-parameter Lamperti PDF \cite{BarkaiJSP06}:
\begin{align}
\label{Lamperti_assym} f({\cal
R},p_+)&=\frac{\sin(\pi\alpha/2)}{\pi}\times
\\ &\times \frac{{\cal
R}(p_+)^{\alpha/2-1}(1-p_+)^{\alpha/2-1}}{(p_+)^{\alpha}+{\cal
R}^2(1-p_+)^{\alpha}+2{\cal
R}(p_+)^{\alpha/2}(1-p_+)^{\alpha/2}\cos(\pi\alpha/2)} \nonumber,
\end{align}
where ${\cal R}$ is the asymmetry parameter. In Eq.
\eqref{Lamperti_eq}, ${\cal R}=1$ as a result of the symmetry of the
walk. Consider, for example, the case when in Eq.
\eqref{occupation_ode} the diffusion coefficient is $\Ka^<$ for
$x<0$ and $\Ka^>$ for $x>0$. Solving the equations as above, we
obtain for $f(p_+)$ the two-parameter Lamperti distribution, Eq.
\eqref{Lamperti_assym}, with ${\cal R}=\sqrt{\Ka^</\Ka^>}$.

Kac proved in 1951 that for $\alpha=1$ (Markovian random-walk), the
occupation time distributions of both Brownian motion and L\'{e}vy
flights obey the same arcsine law \cite{Kac1951}. It was therefore
interesting to find out whether a similar statement holds for
$\alpha<1$. We could not solve the L\'{e}vy flights analog of Eq.
\eqref{occupation_ode}; therefore, we simulated trajectories whose
PDF satisfies the fractional diffusion equation (Eq.
\eqref{fractional_diffusion}) and its generalization to L\'{e}vy
flights (Eq. \eqref{forward_eq_general} with $p=0$). Simulations
were performed using the subordination method described in
\cite{MagdziarzSimulations1,FriedrichSimulations,MagdziarzSimulations2}.
The results are presented in Fig. \ref{fig_occ_levy} and demonstrate
that indeed, for $\alpha<1$, the occupation time distribution is
Lamperti's \eqref{Lamperti_eq} for both $\mu=2$ and $\mu<2$
(L\'{e}vy flights). This result may be related to the recent finding
that the first passage time distribution is also invariant to the
value of $\mu$ \cite{LevySurvivalSimulations}.

\subsection{First passage time}
\label{fpt_section}

The time $t_f$ when a particle starting at $x_0=0$ first hits
$x=b$ is called the first passage time and is a quantity subject to
many studies in physics and other fields \cite{Redner_book}.
The distribution of first passage times for anomalous
paths can be obtained from our fractional Feynman-Kac equation using
an identity due to Kac \cite{Kac1951}:
\begin{equation}
\label{FPT_limit} \textrm{Pr}\{t_f>t\}=\textrm{Pr}\{\max_{0\leq \tau
\leq t}x(\tau)<b\} = \lim_{p\to \infty}G_{x_0}(p,t),
\end{equation}
where the functional is $A_f=\int_0^tU[x(\tau)]d\tau$, and
\begin{equation}
\label{FPT_functional_def} U(x) =
\begin{cases}
0 & x < b, \\
1 & x > b.
\end{cases}
\end{equation}
This is true since
$G_{x_0}(p,t)=\int_0^{\infty}e^{-pA_f}G_{x_0}(A_f,t)dA_f$, and thus,
if the particle has never crossed $x=b$, we have $A_f=0$ and
$e^{-pA_f}=1$, while otherwise, $A_f>0$ and for $p\to \infty$,
$e^{-pA_f}=0$. To find $G_{x_0}(p,t)$ we solve the following
backward equation
\begin{equation*}
sG_{x_0}(p,s)-1 =
\begin{cases}
\Ka s^{1-\alpha}\frac{\di^2}{\di
x_0^2}G_{x_0}(p,s) & x_0<b, \\
\Ka(s+p)^{1-\alpha}\frac{\di^2}{\di
x_0^2}G_{x_0}(p,s)-pG_{x_0}(p,s) & x_0>b.
\end{cases}
\end{equation*}
Solving these equations as in the previous subsection, demanding
that $G_{x_0}(p,s)$ is finite for $|x_0|\to \infty$ and demanding
continuity of $G_{x_0}(p,s)$ and its first derivative at $x_0=b$, we
obtain for $x_0=0$
\begin{equation*}
G_{0}(p,s)=\frac{1}{s}\left[1-e^{-\frac{b}{\sKa}s^{\alpha/2}}\frac{p(s+p)^{\alpha/2-1}}{s^{\alpha/2}+(s+p)^{\alpha/2}}\right].
\end{equation*}
To find the first passage time distribution we take the limit of
infinite $p$,
\begin{equation}
\label{G_FPT_infp} \lim_{p\to \infty}G_{0}(p,s)=
\frac{1}{s}\left(1-e^{-\frac{b}{\sKa}s^{\alpha/2}}\right).
\end{equation}
Defining $\tau_f=(b^2/\Ka)^{1/\alpha}$, we invert $s\to
t$:
\begin{equation*}
\lim_{p\to
\infty}G_{0}(p,t)=\textrm{Pr}\{t_f>t\}=1-\int_0^t\frac{1}{\tau_f}l_{\alpha/2}\left(\frac{\tau}{\tau_f}\right)d\tau,
\end{equation*}
where $\l_{\alpha/2}(t)$ is the one-sided L\'{e}vy distribution of
order $\alpha/2$, whose Laplace transform is
$l_{\alpha/2}(s)=e^{-s^{\alpha/2}}$. The PDF of the first passage
times, $f(t)$, satisfies $f(t)=\frac{\di}{\di
t}\left(\textrm{Pr}\{t_f<t\}\right)=\frac{\di}{\di
t}\left(1-\textrm{Pr}\{t_f>t\}\right)$. Thus,
\begin{equation}
\label{fpt_pdf_levy} f(t) =
\frac{1}{\tau_f}l_{\alpha/2}\left(\frac{t}{\tau_f}\right).
\end{equation}
This result has been previously derived using different methods
(e.g., Eq. (53) of \cite{BarkaiPRE01}). The long times behavior of
$f(t)$ is obtained from the $s\to 0$ limit:
\begin{equation*}
f(s)\sim 1-\frac{b}{\sKa}s^{\alpha/2}.
\end{equation*} Therefore, for long times
\begin{equation}
\label{fpt_long_time_decay}
f(t)\sim
\frac{b}{|\Gamma(-\frac{\alpha}{2})|\sKa}t^{-(1+\alpha/2)}.
\end{equation}
For $\alpha=1$, we reproduce the famous $t^{-3/2}$ decay law of a
one-dimensional random walk \cite{Redner_book}.

\subsection{The maximal displacement}

The maximal displacement of a diffusing particle is a random
variable whose study has been of recent interest (see, e.g.,
\cite{Xmax1,Xmax2,Xmax3,Xmax4} and references therein). To obtain
the distribution of this variable, we use the functional defined in
the previous subsection (Eq. \eqref{FPT_functional_def}). Let
$x_m\equiv \displaystyle\max_{0\leq \tau \leq t}x(\tau)$, and recall
from Eq. \eqref{FPT_limit} that
$\textrm{Pr}\{x_m<b\}=\displaystyle\lim_{p\to \infty}G_{x_0}(p,t)$.
From the previous subsection we have, for $x_0=0$ (Eq.
\eqref{G_FPT_infp})
\begin{equation*}
\textrm{Pr}\{x_m<b\}=\frac{1}{s}\left(1-e^{-\frac{b}{\sqrt{\Ka}}s^{\alpha/2}}\right).
\end{equation*}
Hence, the PDF of $x_m$ is
\begin{equation*}
P(x_m,s)=\frac{s^{\alpha/2-1}}{\sqrt{\Ka}}e^{-\frac{x_m}{\sqrt{\Ka}}s^{\alpha/2}}.
\end{equation*}
Inverting $s\to t$, we obtain
\begin{equation}
P(x_m,t)=\frac{2}{\alpha\sKa}\frac{t}{\left(x_m/\sKa\right)^{1+2/\alpha}}l_{\alpha/2}\left[\frac{t}{\left(x_m/\sKa\right)^{2/\alpha}}\right]\quad;\quad x_m>0.
\end{equation}
This PDF has the same shape as the PDF of $x$ up to a scale factor
of 2 \cite{BarkaiPRE00}, and it is in agreement with the very recent
result of \cite{Xmax3}, derived using a renormalization group method.

\subsection{The hitting probability}

The probability $Q_L(x_0)$ of a particle starting at $0<x_0<L$ to
hit $L$ before hitting $0$ is called the hitting (or exit)
probability. The hitting probability has been investigated long time
ago for Brownian particles \cite{Redner_book} and more recently for
some anomalous processes \cite{HittingMajumdar}. For CTRW, it can be
calculated using the following functional:
\begin{equation}
\label{funct_hitting}
U(x)=\begin{cases}0&0<x<L,\\\infty&\textrm{Otherwise}.\end{cases}
\end{equation}
With Eq. \eqref{funct_hitting}, $A=\int_0^tU[x(\tau)]d\tau=0$ as
long as the particle did not leave the interval $[0,L]$ and is
otherwise infinite. Therefore,
$G(x,p,t)=\int_0^{\infty}e^{-pA}G(x,A,t)dA$ represents the
probability of the particle to be at $x$ at time $t$ without ever
leaving $[0,L]$. This is true for all $p$, since $e^{-pA}$ is either
0 or 1 regardless of $p$. At the boundaries,
$G(x=0,p,t)=G(x=L,p,t)=0$. At $(0,L)$, the forward fractional
Feynman-Kac equation (Eq. \eqref{forward_eq_ps}) reads, in $s$
space,
\begin{equation}
\label{feynman-kac_hitting} sG(x,s)-\delta(x-x_0)=\Ka
s^{1-\alpha}\frac{\di^2}{\di x^2}G(x,s).
\end{equation}
Note that Eq. \eqref{feynman-kac_hitting} does not depend on $p$ and
is equivalent to the fractional diffusion equation, Eq.
\eqref{fractional_diffusion}, with absorbing boundary conditions.
The solution of Eq. \eqref{feynman-kac_hitting} for $x\ne x_0$ is
\begin{equation*}
G(x,s)=\begin{cases}C_0\sinh\left[\frac{s^{\alpha/2}}{\sKa}x\right]&x<x_0,\\
C_1\sinh\left[\frac{s^{\alpha/2}}{\sKa}(L-x)\right]&x>x_0.\end{cases}
\end{equation*}
Matching the solution at $x=x_0$ and demanding $\frac{\di}{\di
x}G(x=x_0^+,s)-\frac{\di}{\di x}G(x=x_0^-,s)=-\frac{1}{\Ka
s^{1-\alpha}}$ (from Eq. \eqref{feynman-kac_hitting}), we have, for
$x>x_0$,
\begin{equation}
\label{sol_Gxs_hitting} G(x,s)=\frac{1}{\sKa
s^{1-\alpha/2}}\frac{\sinh\left(\frac{s^{\alpha/2}}{\sKa}x_0\right)}{\sinh\left(\frac{s^{\alpha/2}}{\sKa}L\right)}\sinh\left[\frac{s^{\alpha/2}}{\sKa}(L-x)\right]\;\;;\;x>x_0.
\end{equation}
The flux of particles that have never before left $[0,L]$ and that
are leaving $[0,L]$ at time $t$ through the right boundary is
\cite{BarkaiPRL99}
\begin{equation*}
\label{flux_def} J(L,t)=-\Ka{\cal
D}_{\textrm{RL},t}^{1-\alpha}\frac{\di}{\di x}G(x=L,t),
\end{equation*}
where ${\cal D}_{\textrm{RL},t}^{1-\alpha}$ is the Riemann-Liouville
fractional derivative, equal to $s^{1-\alpha}$ in Laplace $t\to s$
space (see Eq. \eqref{fractional_diffusion}). The hitting
probability is the sum over all times of the flux through $L$
\cite{Redner_book}:
\begin{equation*}
Q_L(x_0)=\int_0^{\infty}J(L,t)dt=-\Ka
\left.s^{1-\alpha}\frac{\di}{\di x}G(x=L,s)\right|_{s=0}.
\end{equation*}
Using Eq. \eqref{sol_Gxs_hitting}, we have
\begin{equation}
\label{hitting_solution} Q_L(x_0)=\frac{x_0}{L}.
\end{equation}
The hitting probability for anomalous diffusion, $\alpha<1$, is the
same as in the Brownian case \cite{Redner_book}. This is expected,
since the hitting probability should not depend on the waiting time
PDF $\psi(\tau)$.

Note that a backward equation for $Q_L(x_0)$ can be obtained by the
much simpler argument that for unbiased CTRW on a lattice,
$Q_L(x_0)=\left[Q_L(x_0+a)+Q_L(x_0-a)\right]/2$. In the continuum
limit, $a\to 0$, this gives $\frac{\di^2 Q_L(x_0)}{\di x_0^2}=0$.
With the boundary conditions $Q_L(x_0=0)=0$ and $Q_L(x_0=L)=1$, Eq.
\eqref{hitting_solution} immediately follows (see \cite{Redner_book}
for a binomial random walk).

\subsection{The time in an interval}

\label{sect_interval}

Consider the time-in-interval functional
$T_i=\int_0^tU[x(\tau)]d\tau$, where
\begin{equation}
\label{interval_func_def}
U(x)=
\begin{cases}
1 & |x|<b, \\
0 & |x|>b.
\end{cases}
\end{equation}
Namely, $T_i$ is the total residence time of the particle in
the interval $[-b,b]$. Denote by $G_{x_0}(T_i,t)$ the PDF of $T_i$
at time $t$ when the process starts at $x_0$, and denote by $G_{x_0}(p,s)$ the Laplace transform $T_i\to
p$, $t\to s$ of $G_{x_0}(T_i,t)$.
$G_{x_0}(p,s)$ satisfies the backward fractional Feynman-Kac
equation:
\begin{equation}
sG_{x_0}(p,s)-1 =
\begin{cases}
\Ka\left(s+p\right)^{1-\alpha}\frac{\di^2}{\di
x_0^2}G_{x_0}(p,s)-pG_{x_0}(p,s) & |x_0|<b, \\
\Ka s^{1-\alpha}\frac{\di^2}{\di x_0^2}G_{x_0}(p,s) & |x_0|>b.
\end{cases}
\end{equation}
We solve this equation demanding that the solution is finite for
$|x_0| \to \infty$,
\begin{equation}
\label{R_delta} G_{x_0}(p,s)=
\begin{cases}
C_1\cosh\left[x_0(s+p)^{\alpha/2}/\sKa\right]+\frac{1}{s+p} & |x_0|<b, \\
C_0\exp\left[-|x_0|s^{\alpha/2}/\sKa\right]+\frac{1}{s} & |x_0|>b.
\end{cases}
\end{equation}
Demanding continuity of $G_{x_0}(p,s)$ and its first derivative at
$x_0=b$ we solve for $C_1$ and then obtain for $x_0=0$
\begin{equation}
\label{interval_ps}
G_0(p,s)=\frac{p+s\left\{\cosh\left[\frac{(s+p)^{\alpha/2}}{\sKa}b\right]+\frac{(s+p)^{\alpha/2}}{s^{\alpha/2}}\sinh\left[\frac{(s+p)^{\alpha/2}}{\sKa}b\right]\right\}}{s(s+p)\left\{\cosh\left[\frac{(s+p)^{\alpha/2}}{\sKa}b\right]+\frac{(s+p)^{\alpha/2}}{s^{\alpha/2}}\sinh\left[\frac{(s+p)^{\alpha/2}}{\sKa}b\right]\right\}}.
\end{equation}
In principle, the PDF $G_0(T_i,t)$ can be obtained from
\eqref{interval_ps} by inverse Laplace transforming $p\to T_i$ and
$s\to t$. However, we could invert Eq. \eqref{interval_ps} only for $\alpha\to 0$:
\begin{equation}
\label{interval_alpha0}
G_0(T_i,t)_{\alpha\to 0}=(1-e^{-b/\sqrt{K_0}})\delta(T_i-t)+e^{-b/\sqrt{K_0}}\delta(T_i).
\end{equation}
This can be intuitively explained as follows. For $\alpha\to 0$, the
PDF of $x$ becomes time-independent and approaches $G(x,t) \approx
\exp(-|x|/\sqrt{K_0})/(2\sqrt{K_0})$ (Eq. (A1) in
\cite{BarkaiPRE00}). With probability
$\int_{-b}^{b}G(x,t)dx=1-e^{-b/\sqrt{K_0}}$, the particle never
leaves the region $[-b,b]$ and thus $T_i=t$; with probability $e^{-b/\sqrt{K_0}}$, the particle
is almost never at $[-b,b]$ and thus $T_i=0$.

\begin{figure}
\begin{center}
\epsfig{file=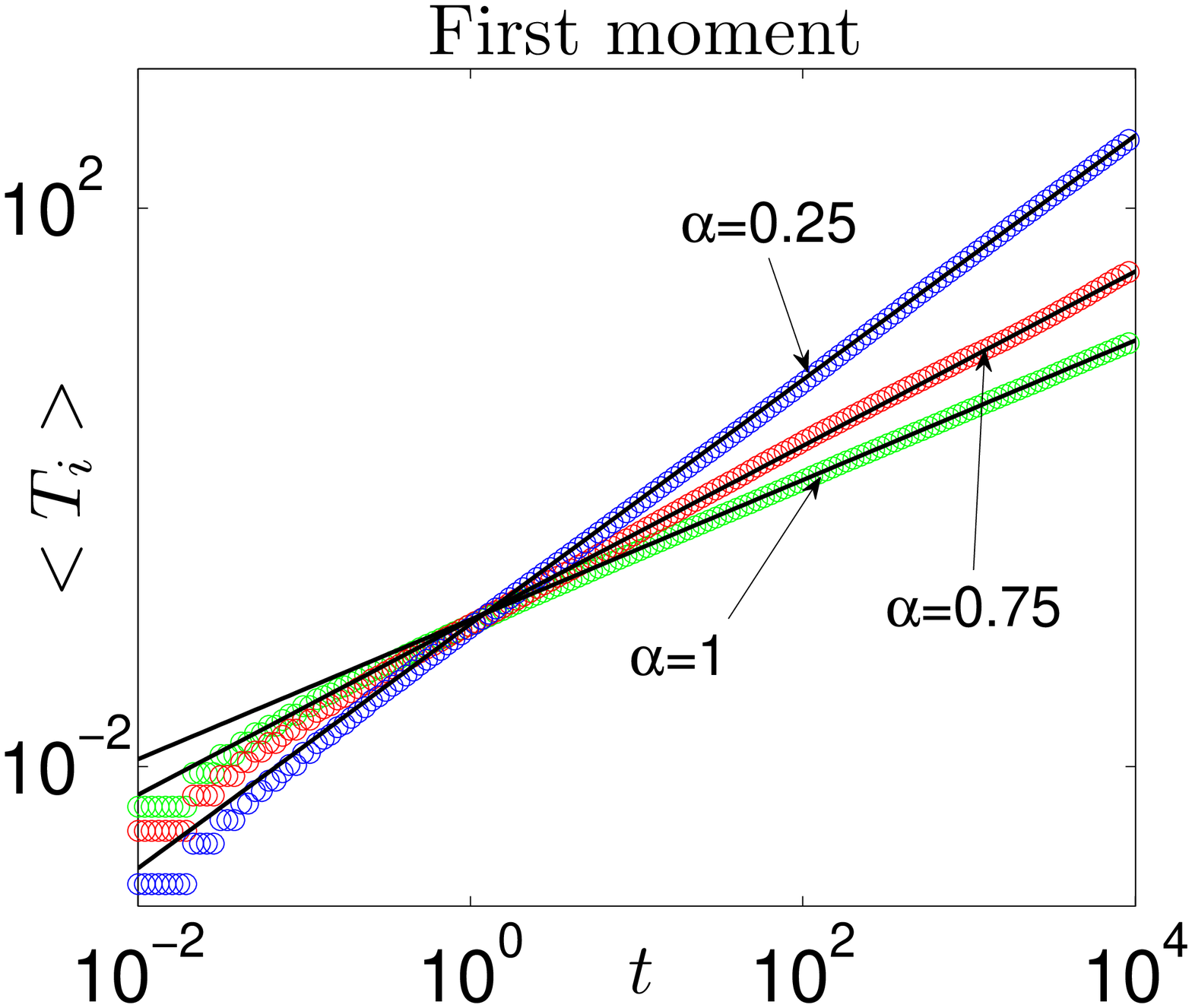,height=7cm,width=8cm}
\epsfig{file=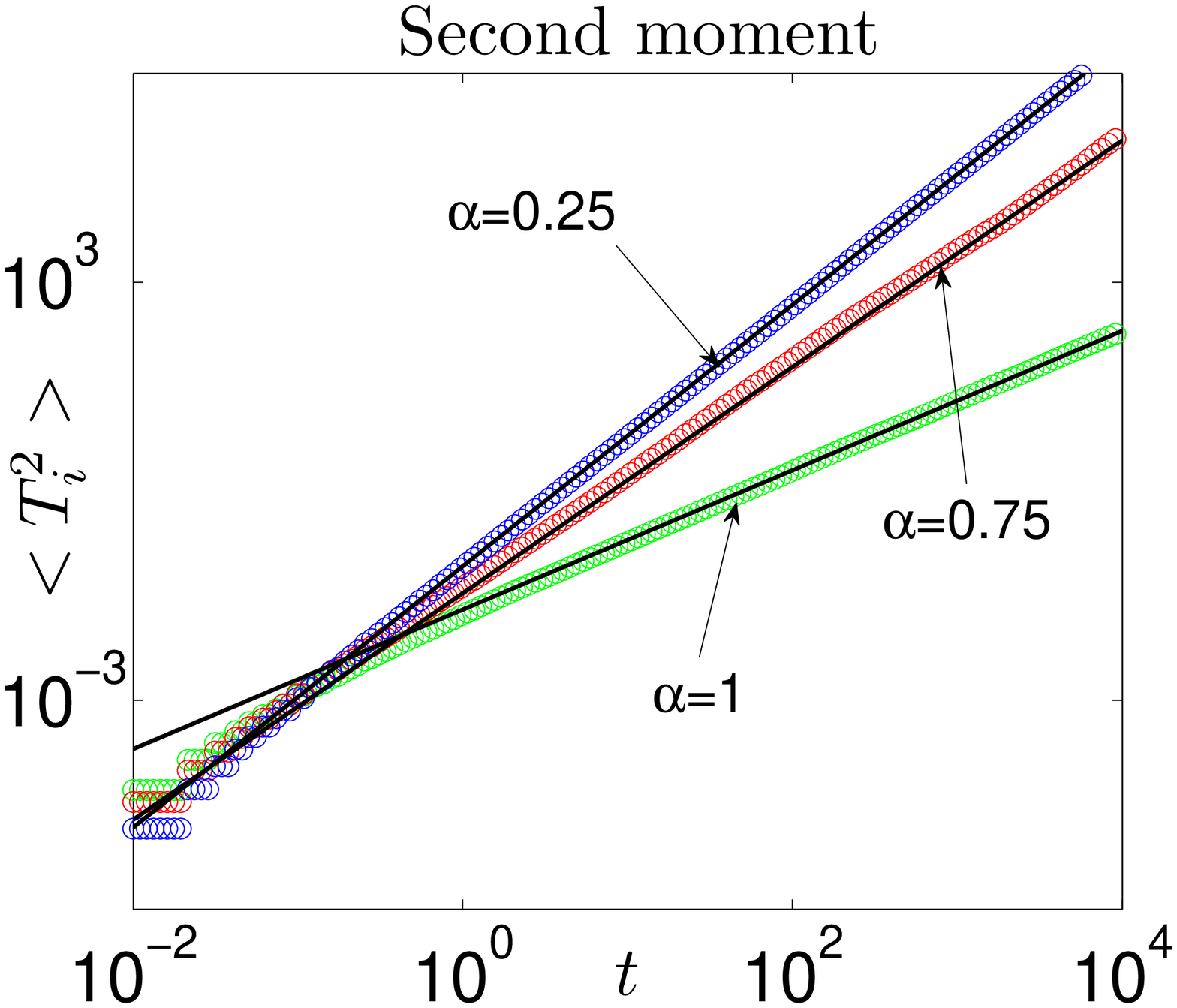,height=7cm,width=8cm}
\end{center} \caption{The first two moments of the time-in-interval
functional $T_i$. Top panel- first moment; bottom panel- second moment.
Simulations (circles) were performed using the method of
\cite{MagdziarzSimulations1,FriedrichSimulations} with parameters as in Fig.
\ref{fig_occ_levy}.
The theoretical curves (lines) correspond to the leading terms of Eq. \eqref{avg_origin}.
The width of the interval (Eq. \eqref{interval_func_def}) is $b=0.1$.}
\label{fig_localtime}
\end{figure}

The first two moments of $T_i$ can be obtained
from Eq. \eqref{interval_ps} by
\begin{equation*}
\av{T_i}(s)=\left.-\frac{\di}{\di p}G_0(p,s)\right|_{p=0}\quad
;\quad\av{T_i^2}(s)= \left.\frac{\di^2}{\di
p^2}G_0(p,s)\right|_{p=0}.
\end{equation*}
Calculating the derivatives, substituting $p=0$, and inverting, we
obtain, in the long times limit,
\begin{align}
\label{avg_origin} \av{T_i}&\sim
t^{1-\alpha/2}\frac{b}{\sKa\Gamma(2-\alpha/2)},\nonumber
\\ \av{T_i^2}&\sim
t^{2-\alpha/2}\frac{2b(1-\alpha)}{\sKa\Gamma(3-\alpha/2)}+t^{2-\alpha}\frac{b^2(3\alpha-1)}{\Ka\Gamma(3-\alpha)}.
\end{align}
We verified that Eq. \eqref{avg_origin} agrees with simulations
(Fig. \ref{fig_localtime}). The average time at $[-b,b]$ scales as
$t^{1-\alpha/2}$ since this is the product of the average number of
returns to the interval $[-b,b]$ ($\sim t^{\alpha/2}$) and the
average time spent at $[-b,b]$ on each visit ($\sim t^{1-\alpha}$;
see Eq. (61) in \cite{BarkaiJSP06}). We also see that for
$\alpha<1$, the PDF of $T_i$ cannot have a scaling form since
$\av{T_i^2}\sim t^{2-\alpha/2}\nsim\av{T_i}^2\sim t^{2-\alpha}$. For
$\alpha=1$, $\av{T_i}\sim t^{1/2}$ and $\av{T_i^2}\sim t$.

\subsection{Survival in a medium with an absorbing interval}
\label{survival_section}

A problem related to that of the previous subsection is a medium in
which a diffusing particle is absorbed at rate $R$ whenever it is in
the interval $[-b,b]$. The survival probability of the particle,
$S$, is related to $T_i$, the total time at $[-b,b]$, through
$S=\exp(-RT_i)$. Thus, if $G_{x_0}(T_i,t)$ is the PDF of $T_i$ at
time $t$, then the $T_i\to R$ Laplace transform
$G_{x_0}(R,t)=\int_0^{\infty}e^{-RT_i}G_{x_0}(T_i,t)dT_i$ equals
$\av{S}$, the survival probability averaged over all trajectories
\cite{GrebenkovPRE}. From Eq. \eqref{interval_ps} of the previous
subsection we immediately obtain (in Laplace $t\to s$ space and for
$x_0=0$)
\begin{equation}
\label{survival_ps}
\av{S}=G_0(R,s)=\frac{R+s\left\{\cosh\left[\frac{(s+R)^{\alpha/2}}{\sKa}b\right]+\frac{(s+R)^{\alpha/2}}{s^{\alpha/2}}\sinh\left[\frac{(s+R)^{\alpha/2}}{\sKa}b\right]\right\}}{s(s+R)\left\{\cosh\left[\frac{(s+R)^{\alpha/2}}{\sKa}b\right]+\frac{(s+R)^{\alpha/2}}{s^{\alpha/2}}\sinh\left[\frac{(s+R)^{\alpha/2}}{\sKa}b\right]\right\}},
\end{equation}
where here $R$ is a parameter (the absorption rate) and thus the equation
needs to be inverted only with respect to $s$. We could invert
\eqref{survival_ps} for a few limiting cases.

\begin{figure}
\begin{center}
\epsfig{file=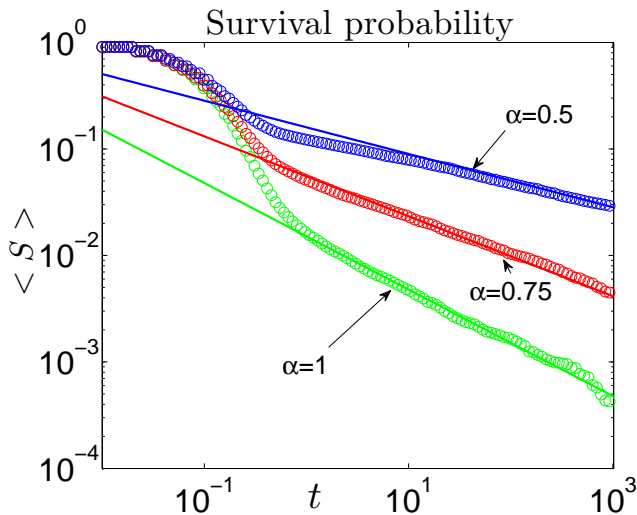,height=7cm,width=9cm}
\end{center} \caption{Survival probability in an absorbing medium.
We simulated anomalous diffusion trajectories as in Fig.
\ref{fig_occ_levy} \cite{MagdziarzSimulations1,FriedrichSimulations}, with total time $t=10^3$.
We plot $\av{e^{-RT_i}}$, where $T_i$ is the total time at
$[-b,b]$, $R=10$, and $b=1$ (Eq. \eqref{interval_func_def}). Simulation results (symbols)
agree with theory (lines), Eq.
\eqref{survival_avg}, for $Rt\gg 1$.} \label{fig_survival}
\end{figure}

\emph{(i)} $t\to \infty$. The long time behavior is obtained by
taking the $s\to 0$ limit and inverting:
\begin{equation}
\label{survival_avg} \av{S}\sim
\left[\Gamma(1-\alpha/2)\sinh\left(\frac{bR^{\alpha/2}}{\sKa}\right)\right]^{-1}(Rt)^{-\alpha/2}+{\cal
O}\left[(Rt)^{-\alpha}\right].
\end{equation}
Thus, the survival probability of the particle in the absorbing
domain decays as $t^{-\alpha/2}$. We verified Eq.
\eqref{survival_avg} using simulations (Fig. \ref{fig_survival}).

\emph{(ii)} $\alpha\to 0$. Inverting Eq. \eqref{survival_ps} yields
\begin{equation}
\label{survival_alpha0}
\av{S}_{\alpha\to 0}=(1-e^{-b/\sqrt{K_0}})e^{-Rt}+e^{-b/\sqrt{K_0}}.
\end{equation}
This can be explained as in the previous subsection. For $\alpha\to
0$, the PDF of $x$ approaches $G(x,t) \approx
\exp(-|x|/\sqrt{K_0})/(2\sqrt{K_0})$. With probability
$\left(1-e^{-b/\sqrt{K_0}}\right)$, the particle never leaves the
region $[-b,b]$. Thus, its probability of survival is just
$e^{-Rt}$. With probability $e^{-b/\sqrt{K_0}}$, the particle is
almost never in the absorbing zone, and it survives with probability
1.

\emph{(iii)} \emph{Other limiting cases.} It can be shown that for $b\to 0$
or $R\to 0$, $\av{S}=1$; for $R\to \infty$, $\av{S}=0$; and for
$b\to \infty$, $\av{S}=e^{-Rt}$.

\subsection{The area under the random walk curve}

\label{sect_area}

The functional $A_x=\int_0^t x(\tau)d\tau$ ($U(x)=x$) represents the
total area under the random walk curve $x(t)$
\cite{FriedrichPLA06,NMRReview}, and it is also related to the phase
accumulated by spins in an NMR experiment \cite{NMRReview}. In this
subsection we obtain the first two moments of this functional, and
for a couple of special cases, also its PDF. Since $A_x$ is not
necessarily positive, we use the generalized forward equation (Eq.
\eqref{forward_eq_fourier} Laplace transformed $t\to s$),
\begin{equation}
\label{forward_eq_x} sG(x,p,s)-\delta(x)=\Ka\frac{\di^2}{\di
x^2}(s-ipx)^{1-\alpha}G(x,p,s)+ipxG(x,p,s).
\end{equation}
Here, $G(x,p,s)$ is the Fourier-Laplace transform of $G(x,A_x,t)$
and we assumed $x_0=0$. Since the walk is unbiased, $\av{A_x}=0$. To
find the second moment of $A_x$, we use
\begin{equation*}
\av{A_x^2}(t)=\int_{-\infty}^{\infty}\left.-\frac{\di^2}{\di
p^2}G(x,p,t)\right|_{p=0}dx.
\end{equation*}
Integrating Eq. \eqref{forward_eq_x} over all $x$, taking the
derivatives with respect to $p$ and substituting $p=0$, we obtain
\begin{equation}
\label{avg_x_A2} s\av{A_x^2}(s)=2\av{xA_x}(s),
\end{equation}
which is in fact obvious since
$\frac{d}{dt}\left(A_x\right)=\frac{d}{dt}\left(\int_0^tx(\tau)d\tau\right)=x$,
and thus $\frac{d}{d t}\av{A_x^2}=2\av{xA_x}$. Hence, the problem of
finding $\av{A_x^2}$ reduces to that of finding $\av{xA_x}$, for
which we have
$\av{xA_x}=\int_{-\infty}^{\infty}-ix\left.\frac{\di}{\di
p}G(x,p,t)\right|_{p=0}dx$. This leads to
\begin{equation}
\label{avg_x_Ax} s\av{xA_x}(s)=\av{x^2}(s).
\end{equation}
Similarly,
\begin{equation}
\label{avg_x_x2} s\av{x^2}(s)=2\Ka s^{-\alpha}.
\end{equation}
Combining Eqs. \eqref{avg_x_A2}, \eqref{avg_x_Ax}, and
\eqref{avg_x_x2}, we find
$\av{A_x^2}(s)=4\Ka s^{-(3+\alpha)}$, or, in $t$ space,
\begin{equation}
\label{area_moment2}
\av{A_x^2}(t)=\frac{4\Ka}{\Gamma(3+\alpha)}t^{2+\alpha}.
\end{equation}
Higher moments of $A_x$ can be similarly calculated (see next
subsection).
The distribution of $A_x$ can be obtained for a few limiting cases.
For $\alpha=1$, $A_x$ is normally distributed (Eq. (61) in
\cite{FriedrichPLA06}):
\begin{equation}
\label{area_alpha1} G(A_x,t)_{\alpha=1}=\sqrt{\frac{3}{4\pi
K_1t^3}}\exp\left(-\frac{3A_x^2}{4K_1t^3}\right).
\end{equation}
For $\alpha\to 0$, the PDF of $x$
is $G(x,t) \approx \exp(-|x|/\sqrt{K_0})/(2\sqrt{K_0})$
(\cite{BarkaiPRE00} and Section \ref{sect_interval})
and is independent of $t$. In other words, the particle is found at
$x(t)$ for most of the time interval $[0,t]$. Hence, $A_x(t)
\approx tx(t)$ and
\begin{equation}
\label{area_alpha0} G(A_x,t)_{\alpha\to 0} \approx
\frac{1}{2\sqrt{K_0}t}\exp\left(-\frac{|A_x|}{\sqrt{K_0}t}\right).
\end{equation}

\begin{figure}
\begin{center}
\hspace{-0.5cm} 
\epsfig{file=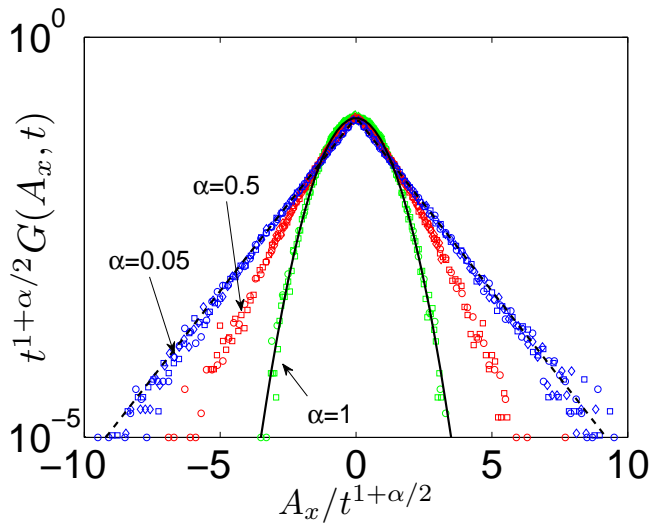,height=7cm,width=9cm} \vspace{0.5cm}
\end{center}
\caption{The area under the random walk curve $A_x$. We simulated
diffusion trajectories as in Fig. \ref{fig_occ_levy}
\cite{MagdziarzSimulations1,FriedrichSimulations} ($10^5$
trajectories) and calculated the PDF $G(A_{x},t)$. To illustrate the
scaling, we plot $t^{1+\alpha/2}G(A_{x},t)$ vs.
$A_{x}/t^{1+\alpha/2}$ ($\Ka=1$), collapsing all curves with the
same $\alpha$ but different times: $t=1$ (circles), $t=10$
(squares), and $t=100$ (diamonds). Theory for $\alpha=1$ is from Eq.
\eqref{area_alpha1} (Gaussian, solid line); theory for $\alpha\to 0$
is from Eq. \eqref{area_alpha0} (exponential, dashed line).}
\label{fig_xk}
\end{figure}

To confirm Eqs. \eqref{area_alpha1} and \eqref{area_alpha0}, we plot
in Fig. \ref{fig_xk} the PDF of $A_x$ for various values of $\alpha$
as obtained from simulation of diffusion trajectories. It can also
be seen from Fig. \ref{fig_xk} that the PDF of $A_x$ obeys a scaling
relation, as we show in the next subsection.


\subsection{The moments of the functionals $U(x)=x^k$}

In the previous subsection we derived the first two moments of the
$U(x)=x$ functional; but in fact, all moments of all functionals
$A_{x^k}=\int_0^tx^k(\tau)d\tau$, $k=1,2,3,...$ can be obtained,
leading to a scaling form of their PDF. As explained above, the
functionals with $k=1,2$ arise in the context of NMR and are
therefore particularly interesting.

We assume $x_0=0$ and consider the forward equation
\eqref{forward_eq_ps} for even $k$'s:
\begin{equation}
\label{forward_moments}
sG(x,p,s)-\delta(x)=\Ka\frac{\di^2}{\di
{x}^2}\left(s+px^k\right)^{1-\alpha}G(x,p,s)-px^kG(x,p,s).
\end{equation}
Here, $G(x,p,s)$ is the double Laplace transform of $G(x,A_{x^k},t)$
since for even $k$'s $A_{x^k}$ is always positive. We are interested
in the moments $\av{A_{x^k}^n}$, $n=0,1,2,...$; however, to find
these, we must first obtain the more general moments $\av{A^nx^m}$,
$n,m=0,1,2,...$. Operating on each term of Eq.
\eqref{forward_moments} with $(-1)^n\frac{\di^n}{\di p^n}$,
substituting $p=0$, multiplying each term by $x^m$, and integrating
over all $x$, Eq. \eqref{forward_moments} becomes
\begin{align}
\label{xk_moments_recursion}
&s\av{A^nx^m}(s)=\delta_{n,0}\delta_{m,0}+H_{n-1}n\av{A^{n-1}x^{m+k}}(s)+ \nonumber \\
&H_{m-2}\Ka m(m-1)\sum_{j=0}^{n}\binom{n}{j}\left[\prod_{l=0}^{j-1}(1-\alpha-l)\right](-1)^js^{1-\alpha-j}\times
\nonumber
\\ & \times\av{A^{n-j}x^{m+jk-2}}(s),
\end{align}
where $\delta_{i,j}$ is Kronecker's delta function--- $\delta_{i,j}$
equals 1 for $i=j$ and equals zero otherwise; and $H_i$ is the
discrete Heaviside function--- $H_i$ equals 1 for $i\geq 0$ and
equals zero otherwise. It can be proved that Eq.
\eqref{xk_moments_recursion} remains true also for odd $k$'s, when
$A_{x^k}$ can be either positive or negative. Eq.
\eqref{xk_moments_recursion} is satisfied by the following choice of
$\av{A^nx^m}$:
\begin{equation}
\label{moments_ansatz} \av{A^nx^m}(s)=
c_{n,m}(k)\Ka^{\frac{m+nk}{2}}s^{-\left(1+n+\frac{m+nk}{2}\alpha\right)},
\end{equation}
for all $n$ and even $m$ when $k$ is even and for even $(n+m)$ when
$k$ is odd. In all other cases $\av{A^nx^m}=0$ due to symmetry. The
$c_{n,m}$'s are $k$-dependent dimensionless constants that satisfy
the following recursion equation:
\begin{align}
\label{cnm_recursion}
&c_{n,m}(k) = \delta_{n,0}\delta_{m,0} + H_{n-1}nc_{n-1,m+k}(k) + \nonumber \\
&H_{m-2}m(m-1)\sum_{j=0}^{n}\binom{n}{j}\left[\prod_{l=0}^{j-1}(1-\alpha-l)\right](-1)^jc_{n-j,m+jk-2}(k),
\end{align}
with initial conditions $c_{0,0}(k)=1$ and $c_{0,1}(k)=0$. The
moments of $A_{x^k}$ are therefore given in $t$ space by
\begin{equation}
\label{all_moments_xk_t}
\av{A_{x^k}^n}(t)=c_{n,0}(k)\frac{\Ka^{\frac{nk}{2}}}{\Gamma\left(1+n+\frac{n\alpha
k}{2}\right)}t^{n\left(1+\frac{\alpha k}{2}\right)}.
\end{equation}
For example, for $k=1$, $\av{A_x}=\av{A_x^3}=0$, $\av{A_x^2}=4\Ka
t^{2+\alpha}/\Gamma(3+\alpha)$ (Eq. \eqref{area_moment2}), and
$\av{A_x^4}=48(\alpha^2+7\alpha+12)\Ka^2
t^{4+2\alpha}/\Gamma(5+2\alpha)$; while for $k=2$,
$\av{A_{x^2}}=2\Ka t^{1+\alpha}/\Gamma(2+\alpha)$ and
$\av{A_{x^2}^2}=(48+8\alpha)\Ka^2 t^{2+2\alpha}/\Gamma(3+2\alpha)$.

Eq. \eqref{all_moments_xk_t} suggests that the PDF of $A_{x^k}$
obeys the scaling relation
\begin{equation}
\label{xk_scaling} G(A_{x^k},t)=\frac{1}{\Ka^{k/2} t^{1+\alpha
k/2}}g_{\alpha,k}\left(\frac{A_{x^k}}{\Ka^{k/2}t^{1+\alpha k/2}}\right),
\end{equation}
where $g_{\alpha,k}(x)$ is a dimensionless scaling function. To
verify the scaling form of Eq. \eqref{xk_scaling}, we plot in Fig.
\ref{fig_xk} simulation results for the PDF of $A_{x}$ ($k=1$) for
$\alpha\approx 0$ and $\alpha=1$ (for which $G(A_x,t)$ is known---
Eqs. \eqref{area_alpha1} and \eqref{area_alpha0} in the previous
subsection), and for an intermediate value, $\alpha=0.5$. In all
cases the simulated PDF satisfies the scaling form
\eqref{xk_scaling}.

\section{Summary and discussion}

Functionals of the path of a Brownian particle have been
investigated in numerous studies since the development of the
Feynman-Kac equation in 1949. However, an analog equation for
functionals of non-Brownian particles has been missing. Here, we
developed such an equation based on the CTRW model with broadly
distributed waiting times. We derived forward and backward equations
(Eqs. \eqref{forward_eq_derive} and \eqref{backward_eq}) and
generalizations to L\'{e}vy flights (Eqs. \eqref{forward_eq_general}
and \eqref{general_backward_eq}). Using the backward equation, we
derived the PDFs of the occupation time in half-space, the first
passage time, and the maximal displacement, and calculated the
average survival probability in an absorbing medium. Using the
forward equation, we calculated the hitting probability and all the
moments of $U(x)=x^k$ functionals.

The fractional Feynman-Kac equation \eqref{forward_eq_derive} can be
obtained from the integer equation \eqref{Feynman-Kac} by insertion
of a substantial fractional derivative operator
\cite{FriedrichPRL06}. In that sense, our work is a natural
generalization of that of Kac's. The distributions we obtained for
specific functionals are also the expected extensions of their
Brownian counterparts: the arcsine law for the occupation time in
half-space \cite{MajumdarReview,Watanabe} was replaced by Lamperti's
PDF (Eq. \eqref{Lamperti_eq}) \cite{Lamperti}, and the famous
$t^{-3/2}$ decay of the one-dimensional first passage time PDF
\cite{Redner_book} became $t^{-(1+\alpha/2)}$ (Eq.
\eqref{fpt_long_time_decay}). Thus, our analysis supports the notion
that CTRW and the emerging fractional paths
\cite{MagdziarzSimulations1,FriedrichSimulations} are elegant
generalizations of ordinary Brownian motion. Nevertheless, other
non-Brownian processes are also important. For example, it would be
interesting to find an equation for the PDF of anomalous functionals
when the underlying process is fractional Brownian motion
\cite{Mandelbrot}.

Our fractional Feynman-Kac equation \eqref{forward_eq_derive} has
the form of a fractional Schr\"{o}dinger equation in imaginary time.
Real time, fractional Schr\"{o}dinger equations for the wave
function have also been recently proposed
\cite{HuKallianpur,Laskin2000PRE,Laskin2002,Naber,SpaceTimeFSE}.
However, these are very different from our fractional Feynman-Kac
equation. In \cite{HuKallianpur,Laskin2000PRE,Laskin2002}, the
Laplacian was replaced with a fractional spatial derivative which
would correspond to a \emph{Markovian} CTRW with heavy tailed
distribution of jump lengths (L\'{e}vy flights; see also the
Appendix below). The approach in \cite{Naber,SpaceTimeFSE} is based
on a temporal fractional Riemann-Liouville derivative--- however not
substantial--- which leads to non-Hermitian evolution and hence
non-normalizable quantum mechanics. It is unclear yet whether all
these fractional Schr\"{o}dinger equations actually describe any
physical phenomenon (see \cite{Iomin} for discussion). In principle,
a fractional Schr\"{o}dinger equation can also be written using the
substantial fractional derivative we used here. If there is a
physical process behind such a quantum mechanical analog of our
equation remains at this stage unclear.

In this paper we considered only the case of a free particle. In
\cite{BarkaiPRL09}, we reported a fractional Feynman-Kac equation
for a particle under the influence of a binding force, where
anomalous diffusion can lead to weak ergodicity breaking
\cite{BarkaiPRL07,BarkaiPRL05,BarkaiJSP08}. The derivation of an
equation for the distribution of general functionals and the
treatment of specific functionals for bounded particles will be
published elsewhere.

\begin{acknowledgements}
We thank S. Burov for discussions and the Israel Science Foundation
for financial support. S. C. is supported by the Adams Fellowship
Program of the Israel Academy of Sciences and Humanities.
\end{acknowledgements}

\section*{Appendix: Generalization to arbitrary functionals and L\'{e}vy flights}

Here we generalize our forward and backward fractional Feynman-Kac equations
(\eqref{forward_eq_derive} and \eqref{backward_eq}, respectively) to the case
when the functional is not necessarily positive and to the case when the CTRW
jump length distribution is arbitrary, and in particular, heavy tailed.

In our generalized CTRW model, the particle moves, after waiting at
$x$, to $x+\Delta_x$, where $\Delta_x$ is distributed according to
$f(\Delta_x)$. The PDF $f(\Delta_x)$ must be symmetric:
$f(\Delta_x)=f(-\Delta_x)$ but can be otherwise arbitrary. Let us
rederive the forward equation for this model. We replace Eq.
\eqref{Q_recursion} with
\begin{align*}
Q_{n+1}(x,A,t)=&\int_0^t\psi(\tau)\int_{-\infty}^{\infty}f(\Delta_x)Q_n[x-\Delta_x,A-\tau
U(x-\Delta_x),t-\tau]d\Delta_x d\tau.
\end{align*}
Since $A$ can be negative, we \emph{Fourier} transform the last
equation $A\to p$
\begin{equation*}
Q_{n+1}(x,p,t)=\int_0^t\psi(\tau)\int_{-\infty}^{\infty}f(\Delta_x)e^{ip\tau
U(x-\Delta_x)}Q_n(x-\Delta_x,p,t-\tau)d\Delta_x d\tau.
\end{equation*}
Laplace transforming $t\to s$ and Fourier transforming $x\to k$ we
have
\begin{equation*}
Q_{n+1}(k,p,s)=\int_{-\infty}^{\infty}e^{ikx}\int_{-\infty}^{\infty}f(\Delta_x)\hat\psi[s-ipU(x-\Delta_x)]Q_n(x-\Delta_x,p,s)d\Delta_x
dx.
\end{equation*}
Changing variables: $x'=x-\Delta_x$,
\begin{align*}
Q_{n+1}(k,p,s)&=\int_{-\infty}^{\infty}e^{ik\Delta_x}f(\Delta_x)d\Delta_x\int_{-\infty}^{\infty}e^{ikx'}\hat\psi[s-ipU(x')]Q_n(x',p,s)dx'
\nonumber \\ &= f(k)\hat\psi\left[s-ipU\left(-i\frac{\di}{\di
k}\right)\right]Q_n(k,p,s).
\end{align*}
Summing over all $n$ and using the initial condition
$Q_0(k,p,s)=e^{ikx_0}$,
\begin{equation*}
\sum_{n=0}^{\infty}Q_n(k,p,s)=\left\{1-f(k)\hat\psi\left[s-ipU\left(-i\frac{\di}{\di
k}\right)\right]\right\}^{-1}e^{ikx_0}.
\end{equation*}
Note that this agrees with Eq. \eqref{Q_solution} since for nearest
neighbor hopping
$f(k)=\int_{-\infty}^{\infty}e^{ik\Delta_x}\left[\frac{1}{2}\delta(\Delta_x-a)+\frac{1}{2}\delta(\Delta_x+a)\right]d\Delta_x=\cos(ka)$.
Next, we observe that Eq. \eqref{G_Q} of Section
\ref{forward_section} remains the same even under the general
conditions. Calculating the transformed $G(k,p,s)$ as above, and
using the result of the last equation, we obtain the formal solution
\begin{align*}
G(k,p,s)=& \frac{1-\hat\psi\left[s-ipU\left(-i\frac{\di}{\di
k}\right)\right]}{s-ipU\left(-i\frac{\di}{\di k}\right)}\times
\nonumber
\\ & \times\left\{1-f(k)\hat\psi\left[s-ipU\left(-i\frac{\di}{\di
k}\right)\right]\right\}^{-1}e^{ikx_0}.
\end{align*}
We now assume that $f(\Delta_x)$ has a finite second moment and thus
its characteristic function can be written, for small $k$, as $f(k)
\sim 1-\sigma^2k^2/2$. This characteristic function is identical to that of nearest
neighbor hopping (with $\sigma=a$);
we can thus proceed as in Section \ref{forward_section} to obtain
\begin{equation}
\label{forward_eq_fourier_appendix} \frac{\di}{\di
t}G(x,p,t)=\Ka\frac{\di^2}{\di x^2}{\cal
D}_t^{1-\alpha}G(x,p,t)+ipU(x)G(x,p,t),
\end{equation}
where here ${\cal D}_t^{1-\alpha}\to[s-ipU(x)]^{1-\alpha}$ in Laplace $s$ space and
$\Ka=\sigma^2/(2B_{\alpha})$.

Consider now the case of \emph{L\'{e}vy flights}--- $f(\Delta_x)\sim
|\Delta_x|^{-(1+\mu)}$ (for large $\Delta_x$) with $0<\mu<2$, and
thus jump lengths have a diverging second moment. The characteristic
function is $f(k) \sim 1-C_{\mu}|k|^{\mu}$, and the fractional
Feynman-Kac equation becomes
\begin{equation}
\label{forward_eq_levy_appendix} \frac{\di}{\di
t}G(x,p,t)=K_{\alpha,\mu}\nabla_x^{\mu}{\cal
D}_t^{1-\alpha}G(x,p,t)+ipU(x)G(x,p,t),
\end{equation}
where $K_{\alpha,\mu}=C_{\mu}/B_{\alpha}$ and $\nabla_x^{\mu}$ is
the Riesz spatial fractional derivative operator:
$\nabla_x^{\mu}\to -|k|^{\mu}$ in Fourier $k$ space.

Repeating the calculations of Section \ref{backward_section}
for a non-necessarily-positive functional and for L\'{e}vy flights,
it can be shown that the generalized backward equation is:
\begin{equation}
\label{general_backward_eq_appendix} \frac{\di}{\di t}G_{x_0}(p,t) =
K_{\alpha,\mu}{\cal
D}_t^{1-\alpha}\nabla_{x_0}^{\mu}G_{x_0}(p,t)+ipU(x_0)G_{x_0}(p,t).
\end{equation}
Here, ${\cal D}_t^{1-\alpha}\to[s-ipU(x_0)]^{1-\alpha}$ in Laplace $s$ space
and $\nabla_{x_0}^{\mu}\to -|k_0|^{\mu}$ in Fourier $k_0$ space.

\bibliographystyle{unsrt}
\bibliography{ctb}   

\end{document}